\documentclass[a4paper,fleqn,usenatbib,letters]{mnras}

\usepackage[T1]{fontenc}
\usepackage{ae,aecompl}

\usepackage{graphicx}	
\usepackage{amsmath}	
\usepackage{amssymb}	

\usepackage{color}
\usepackage{float}
\usepackage{txfonts}
\usepackage{units}
\usepackage{url}

\newcommand{\fermilat}{\textit{Fermi}-LAT}
\newcommand{\lsi}{LS~I~+61$^{\circ}$303}
\newcommand{\grs}{GRS~1915+105}
\newcommand{\vcyg}{V404~Cyg}
\newcommand{\cgx}{Cyg~X-3}

\title[Radio QPO in \lsi{}]{Radio QPO in the $\gamma$-ray-loud X-ray binary \lsi{}}

\author[F. Jaron et al.]{F.~Jaron,$^{1,2}$\thanks{E-mail: fjaron@mpifr-bonn.mpg.de}
 R.~Sharma,$^{1}$
 M.~Massi,$^{1}$
 L.~Fuhrmann,$^{1,3}$
 E.~Angelakis,$^{1}$
 I.~Myserlis,$^{1}$
\newauthor
 Guang-Xing~Li$^{4}$ and
Xun~Shi$^{5}$
\\
$^{1}$Max-Planck-Institut f\"ur Radioastronomie, Auf dem H\"ugel 69, D-53121 Bonn, Germany \\
$^{2}$Institut f\"ur Geod\"asie und Geoinformation -- IGG, Universit\"at Bonn, Nu\ss allee 17, D-53115 Bonn, Germany\\
$^{3}$Universit\"at Siegen,
  Lehrstuhl f\"ur Hochfrequenzsensoren und Radarverfahren,
  Paul-Bonatz-Str. 9-11
  D-57076 Siegen, Germany\\
$^{4}$Ludwig-Maximilians Universit\"at M\"unchen,
  Department f\"ur Physik,
  Scheinerstr.~1,
  D-81679 M\"unchen, Germany\\
$^{5}$Max-Planck-Institut f\"ur Astrophysik,
  Karl-Schwarzschild-Str.~1,
  D-85748 Garching, Germany\\
}

\date{Accepted XXX. Received YYY; in original form ZZZ}

\pubyear{2017}

\begin{document}
\label{firstpage}
\pagerange{\pageref{firstpage}--\pageref{lastpage}}
\maketitle

\begin{abstract}

\lsi{} is a $\gamma$-ray emitting X-ray binary with periodic radio outbursts with time scales of one month. Previous observations have revealed microflares superimposed on these large outbursts with periods ranging from a few minutes to hours. This makes \lsi{}, along with Cyg~X-1, the only TeV emitting X-ray binary exhibiting radio microflares. To further investigate these microflaring activity in \lsi{}
we observed 
the source with the 100-m Effelsberg radio telescope at 4.85, 8.35, and 10.45~GHz and performed timing analysis on the obtained data. Radio oscillations of 15~hours time scales are detected at all three frequencies. We also compare the spectral index evolution of radio data to that of the photon index of GeV data observed by \fermilat{}. We conclude that the observed QPO could result from multiple shocks in a jet.

\end{abstract}





\section{Introduction}

\begin{figure}
\begin{center}
\includegraphics[width=0.65\linewidth]{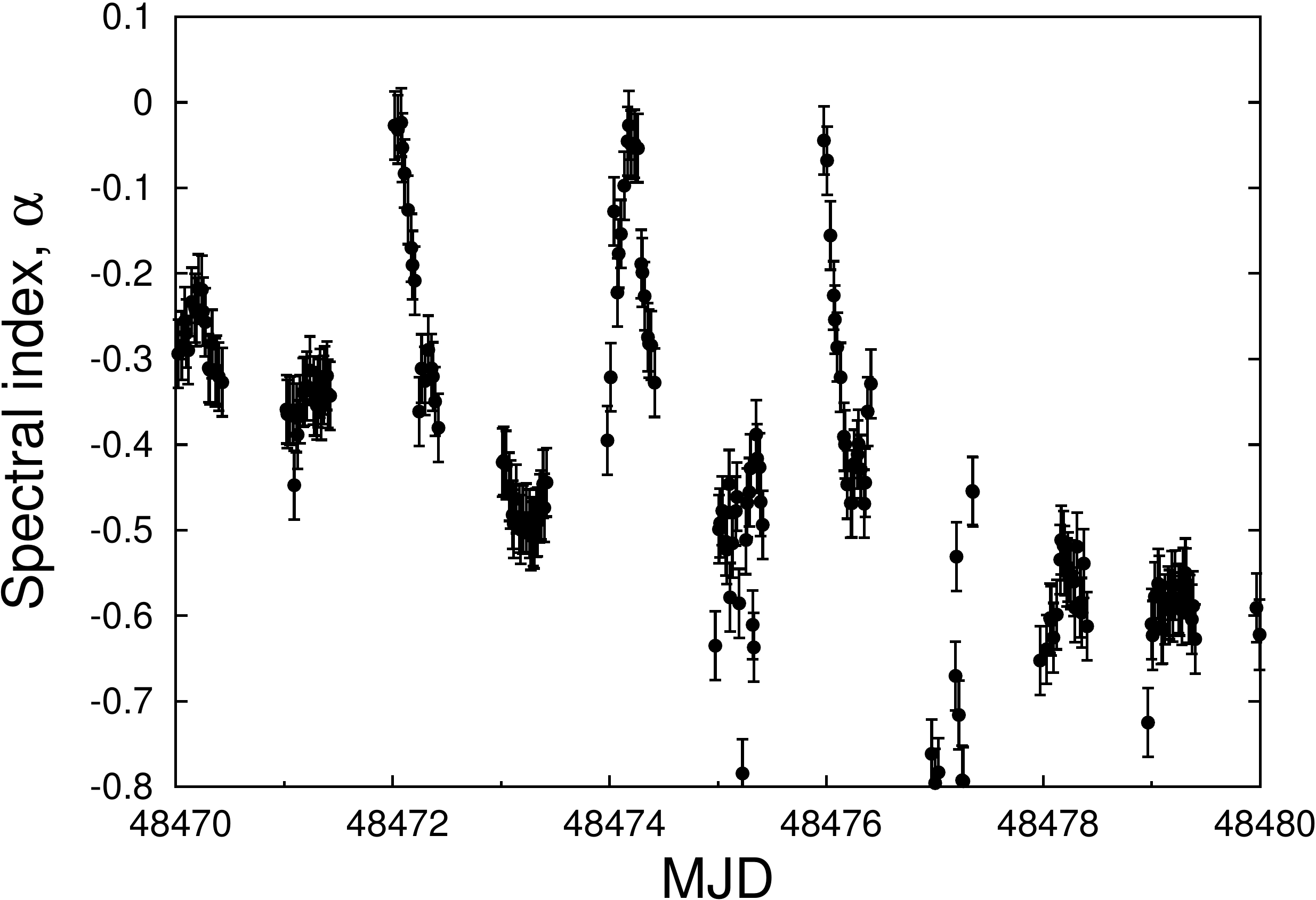}
\caption{Spectral index of Cyg X-3 from Green Bank Interferometer data at 2 and 8~GHz. Three oscillations separated by 3~days are evident. Similar oscillations are present in the flux density at the two frequencies \citep[see][]{Zimmermann2015}.}
\label{fig:cyg}
\end{center}
\end{figure}

It is a known fact that a subclass of X-ray binaries and a subclass of Active Galactic Nuclei (AGN) are sources of radio emission.
There is evidence that radio outbursts in these systems are superimposed by microflaring activity of lower amplitude and short time scales.
These short timescale variations, characterised as Quasi Periodic Oscillations (QPO) \citep{fender97}, may change period between different epochs or are present in a relatively short interval of time, with few oscillations. 
In X-ray binaries with radio jets, i.e., microquasars \citep{Mirabel1999}, QPO were first observed in the black hole system \vcyg{} where a range of 22--120~min sinusoidal radio variations were observed 
during the decay of a radio outburst \citep{Han1992}, and more recently \citet{Plotkin2017} found correlation between radio and X-ray variability on minute time scales.
Short-term radio variability on time scales of $\sim$1~hour was observed in Cyg~X-1 \citep{Marti2001}.
QPO in \grs{} have been extensively studied, e.g., by
\citet{Pooley1997}, who revealed QPO with periods in the range of 20--40~min; observations by \citet{Rodriguez1997} showed oscillations of 30~min. Also, \citet{Fender2002} performed an analysis 
at dual radio frequencies and again revealed QPO with periods of
$\sim$30~min following the decay of a major outburst. Moreover,
the radio spectral index $\alpha$ oscillates from negative to positive values (see Fig.~2 in \citealt{Fender2002}).
\citet{Fender1998} demonstrated that infrared oscillations precede the radio ones, and both oscillations have similar period and shape.
In simultaneous radio and X-ray observations of \grs{}, \citet{Klein-Wolt2002} showed that X-ray dips were associated with radio peaks.
Oscillations with time scales of days are known to occur in 
\cgx{} \citep{Zimmermann2015}. 
These longer period QPO are present in both total flux density and
radio spectral index (Fig.~\ref{fig:cyg}).

QPO seem to be a general phenomenon associated to the ejections in accreting systems and have in fact been observed also in AGN. X-ray QPO of 55~minutes have been indicated in the flat spectrum radio quasar 3C~273 \citep{espaillat2008} and QPO of $\sim$60~minutes have been observed in the narrow-line Seyfert~1 Galaxy\ RE~J1034+396 \citep{Gierlinski2008}. \citet{Rani2010} found evidence for optical QPO of 15~minutes in the blazar S5~0716+714 in $R$-band.

The physical process behind QPO is still a matter of debate, three explanations are discussed.
The first hypothesis explains QPO as the result of a single shock propagating 
down a helical jet and producing increased flux each time the shock meets another twist of the helix at the angle that provides the maximum Doppler boosting for the observer \citep{Rani2010}.
In the second scenario, QPO could be related to shocks passing down the jet and accelerating particles in situ \citep[][and references therein]{Klein-Wolt2002}. 
%
As a third explanation, QPO have been related to discrete ejections of plasma. Multi-wavelength QPO observations of \grs{} (see references in \citealt{Mirabel1999}) have been interpreted as periodic discrete ejections of plasma, with a mass of about $\sim 10^{19}$\,g and at relativistic speeds, with subsequent replenishment of the inner accretion disk.

Testing possible models for the origin of QPO is still complicated because of insufficient statistics. Remaining open questions include: How stable are these ``quasi'' periodic oscillations? Why are QPO present in the radio spectral index? Why does the radio spectrum oscillate between optically thin and thick emission in \grs{} and \cgx{}, and is that a general property of QPO? Do the ``flat'' radio spectra in microquasars and AGN indeed arise from the combination of emission from optically thick and thin regions as suggested in \citet{Fender2002}? In order to answer these questions, it is requisite to increase the sample of X-ray binaries exhibiting radio QPO. For this purpose we investigate on \lsi{}
which is one of the few radio emitting X-ray binaries which also emits in $\gamma$-rays (GeV, \citealt{Abdo2009}, and TeV, \citealt{Albert2006}, \citealt{Acciari2008}). 
Moreover, \lsi{} is periodic at all wavelengths on the orbital time scale (26.5 d) 
and has revealed microflaring activity (Sect.~2).

Aimed at investigating QPO in \lsi{} we performed new observations with the Effelsberg 100-m radio telescope. Our new radio observations and data analysis are described in Sect.~3 along with the description of \fermilat{} data reduction. In Sect.~4 we present our results and in Sect.~5 our conclusions.


\section{The binary system \lsi{}}


The X-ray binary \lsi{} is composed of a Be star \citep{Casares2005} and a black hole \citep{Massi2017}, exhibiting radio outbursts \citep{Gregory2002,Jaron2013} ocurring periodic, related to the orbital period of $\sim 26.5$\,days. 
With an SED peaking above 1\,GeV, \lsi{} can further be classified as a $\gamma$-ray binary (see Table~2 in \citealt{Dubus2013}).
Among the black hole binaries discussed above, only \lsi{} and Cyg~X-1 have the peculiarity of emitting at TeV, 
this, however, being a transient phenomenon for Cyg~X-1 \citep{Albert2006}, and also \lsi{} has episodes of TeV non-detection \citep{Acciari2011}.

Two magnetar-like signals were detected from a large region of sky
crowded by other potential sources beside \lsi{} \citep{Rea2008}. On that basis \citet{Torres2012} put forward their working
hypothesis that \lsi{} could be the first magnetar detected in
a binary system, and studied the implications. In the magnetar
scenario, as well as for the pulsar scenario \citep[see, e.g.,][]{Dubus2013}, variability of the Be
star would trigger and explain observed long-term variability in the emission
of \lsi{} at all wavelengths \citep{Gregory2002,Li2014,Ackermann2013,Ahnen2016}. However, following the well-studied (i.e., over 100 years) case of the binary system $\zeta$~Tau, i.e., also a Be star in a binary system (Sect.~4 in \citealt{MassiTorricelli2016} and references therein), Be star variations last 2-3
cycles only and are of different lengths \citep{Rivinius2013}. Timing analysis of 37 years of radio data performed by \citet{MassiTorricelli2016} reveals that \lsi{} does not show merely
2-3 cycles which are of different lengths but a repetition of 8 full cycles of an
identical length of 1628\,days, well in agreement with the scenario of a microquasar with a precessing jet \citep{Massi2014, Jaron2016}. In addition, as discussed in \citet{Jaron2016}, the orbital shift
in the equivalent width of the H$\alpha$
emission line \citep{Paredes-Fortuny2015},
points to  variations caused by  a  precessing  jet.

Optical polarization observations determined a rotational axis for the Be star of 25~degrees \citep{Nagae2008}. For parallel orbital and Be spin axes and the mass function determined through orbital motions measurements \citep{Casares2005} a 25~degree inclination implies a black hole of 4~solar
masses, as argued by \citet[][and references therein]{Massi2017}, who further probe from X-ray observations that the photon index vs. luminosity trend of \lsi{} is very different from that of the non-accreting pulsar binary PSR B1259-63, 
whereas its trend agrees with that of moderate-luminosity regime black holes in general and with the two black holes in the same X-ray luminosity range: Swift J1357.2-0933 and V404 Cygni, in particular.


%

The source is well-known for its strong radio outbursts with orbital periodicity 
monitored for $\sim 40$~years \citep{MassiTorricelli2016}. Along with this strong radio outburst there is also evidence of occurrance of microflaring activity with time scales of minutes to hours. 
During the decaying phase of one of the radio outbursts of \lsi{}, a step-like pattern was observed for the first time with the Westerbork Synthesis Radio Telescope (WSRT) with a characteristic timescale of $\sim 10^3$\,s \citep{Taylor1992}. \citet{Peracaula1997} performed the first systematic study of this short-term radio variability and found a period of 1.4~hours for these microflares with an amplitude of $\sim$4\,mJy in VLA observations related to the decay of one radio outburst.
Recently, \citet{Zimmermann2015} observed during the decay of one outburst of \lsi{} sub-flares with a characteristic time-scale of two days (see their Fig.~1). Furthermore, the radio spectral index oscillated
in a quasi-regular fashion and the local peaks in spectral index roughly coincide with the peaks of the sub-flares seen in the total intensity light curves. Finally, at higher energy, \citet{Harrison2000} discovered a periodicity of $\sim$40~minutes in an X-ray \textit{ASCA} observation associated to the onset of a radio outburst.
%

\section{Observation and Data analysis}

\setcounter{table}{0}
\begin{table}
\caption{Best-fit parameters of the parabola used for subtracting the long-term trend from the lightcurves.}
\centering
\begin{tabular}{lccc}
\hline
Frequency & $a_1$ & $t_0$ (MJD) & $b_1$   \\ \hline 
4.85 GHz & $-0.009 \pm 0.001$ & 56766.6 $\pm$ 0.1 & 0.16 $\pm$ 0.02  \\
8.35 GHz &  $-0.008 \pm 0.001$ & 56766.3 $\pm$ 0.1 & 0.13 $\pm$ 0.05\\
10.45 GHz & $-0.009 \pm 0.001$ & 56766.4 $\pm$ 0.1 & 0.12 $\pm$ 0.03 \\ \hline
\end{tabular}
\label{table:paramlc}
\end{table}

\subsection{Effelsberg radio telescope}

\begin{figure*}
\begin{center}
\includegraphics{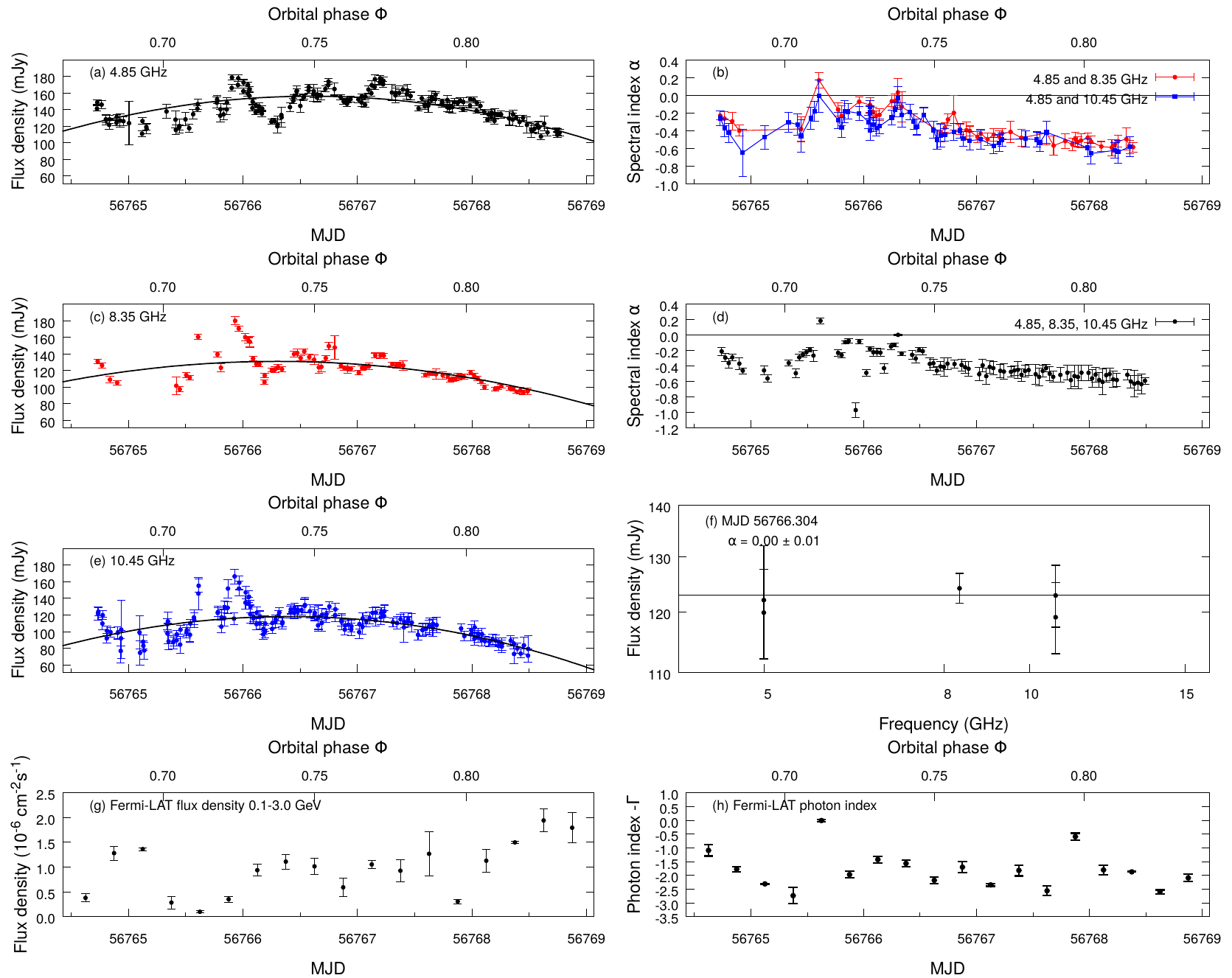}
\caption{(a), (c), (e): Original light curves at indicated frequencies. The light curves were fitted with a parabola (black solid line) in order to remove the long-term trend. (b) Spectral index between indicated frequencies. (d): Spectral index resulting from fitting a power law to all three frequencies. (f): Zoom-in of the flat spectral index at MJD 56766.304, axes in logarithmic scale.  (g): \fermilat{} light curve. (h): \fermilat{} photon index.}
\label{fig:lightcurves}
\end{center}
\end{figure*}

Our multi-frequency flux density measurements  
were performed about every 45 minutes for almost 100~hours on April 17$-$21, 2014 (MJD 56764.726 until MJD~56768.763). The orbital phase $\Phi$ is defined as 
$
\Phi = \dfrac{t-t_0}{P_{\textrm{orb}}} - \textrm{int} \left(\dfrac{t-t_0}{P_\textrm{{orb}}}\right),
$
where, $t_0$ = MJD 43366.275, orbital period $P_{\textrm{orb}}=26.4960\pm0.0028$ d \citep{Gregory2002}, giving $\Phi=0.68-0.83~$ for our observations. The secondary focus receivers of the Effelsberg 100-m telescope were used
at three frequencies, namely  4.85, 8.35, and 10.45~GHz (6.0, 3.6, 2.8~cm wavelengths respectively).
Flux density measurements were performed using the ``cross-scan'' technique, i.e., progressively
slewing over the source position in azimuthal and elevation direction with the number of sub-scans matching the source brightness at a given frequency (typically 4 to 12). 
At 4.85 and 10.45 \,GHz, the ``beam switch'', realised through multiple-feed systems, removed most of the tropospheric variations, allowing for more accurate measurements. The cross-scan technique on the other hand allows instantaneous correction of small, remaining pointing offsets.
The data reduction, from raw telescope data to calibrated flux densities/spectra, was done in the standard manner as described in \citet{Angelakis2015}.
Problems with the 8.35~GHz receiver caused the large flagging of data at this frequency. 
The best data set for sampling rate and SNR is that at 4.85 GHz.
The light curves obtained for all three frequencies are shown in Figure~\ref{fig:lightcurves} (a, c, e) along with their spectral index computed as
$
\alpha = \log(S_1/S_2)/\log(\nu_1/\nu_2)
$
for Fig~\ref{fig:lightcurves}\,b, and as a linear fit to the fluxes vs.~frequency in double logarithmic scale for every time bin of 45~minutes, shown in Fig.~\ref{fig:lightcurves}\,d.

In order to analyse short-term periodicities, we removed the long-term trend from the light curves by subtracting a quadratic function 
\begin{equation}
f_1(t)=a_1 (t-t_0)^2 + b_1,
\end{equation}
with best-fit parameters listed in Table~\ref{table:paramlc}.
The rectified data were then analyzed using wavelet analysis \citep{Torrence1998}, auto-correlation function and Lomb-Scargle periodogram \citep{Lomb1976, Scargle1982}.
We test the significance of found periodic signals by employing the Fisher randomisation test \citep{Nemec1985} where the flux is permuted a thousand times and thousand new randomised time series are created and their periodograms calculacted. The proportion of randomised time series that contain a higher peak in the periodogram than the original periodogram at any frequency then gives the false alarm probability $p$ of the peak. If $p < 0.01$, the period is significant, and if $0.01 < p < 0.1$ the period is marginally significant. 
The data were then folded on the resulting significant period, and the folded data were fitted with a sine-function 
\begin{equation}
f_2(\phi)=A\sin2\pi(\phi-\phi_0)+B,
\end{equation}
with its best-fit parameters in Table \ref{table:paramsine}.    

\subsection{\fermilat{} data reduction}

We compare our radio data with GeV $\gamma$-ray data. For this purpose we use the GeV data from the \fermilat{} in the energy range 0.1--3.0~GeV from MJD~56764.125 until MJD~56768.875. For the analysis of \fermilat{} data we used version v10r0p5 of the \textit{Fermi} ScienceTools\footnote{available from \url{http://fermi.gsfc.nasa.gov/ssc/data/analysis/software/}}. We used the instrument response function P8R2\_SOURCE\_V6 and the corresponding model gll\_iem\_v06.fits for the Galactic diffuse emission and the template iso\_P8R2\_SOURCE\_V6\_v06.txt. Model files were created automatically with the script \texttt{make3FGLxml.py}\footnote{available from \url{http://fermi.gsfc.nasa.gov/ssc/data/analysis/user/}} from the third \fermilat{} point source catalog \citep{FermiLAT2015}. The spectral shape of \lsi{} in the GeV regime is a power law with an exponential cut-off at 4--6~GeV \citep{Abdo2009, Hadasch2012}. Here we restrict our analysis to the power law part of the GeV emission by fitting the source with
\begin{equation}
  \frac{{\rm d}n}{{\rm d}E} = n_0\left(\frac{E}{E_0}\right)^{-\Gamma}
   ~~\left[{\rm counts \over cm^2
       sec~ dE}\right]
       \label{eq:7}
\end{equation}
with all parameters left free for the fit, and including data in the energy range $E = \unit[0.1 - 3]{GeV}$. All other sources within a radius of $10^{\circ}$ and the Galactic diffuse emission were left free for the fit. All sources between $10^{\circ}-15^{\circ}$ were fixed to their catalog values. The light curves 
were computed performing this fit for every time bin  of width half a day for data from 2014~April~17 (MJD~56764.125) till 2014~April~21 (MJD~56768.875). On average, the test statistic for \lsi{} was 40, which corresponds to a detection of the source at the $6.3\sigma$ level on average in each time bin.

\vspace{-12pt}
\section{Results}

\setcounter{table}{1}
\begin{table}
\caption{Best-fit parameters as a result of fitting a sine function to the folded data with period of 15.4 h.} 
\centering
\begin{tabular}{rrrrr}
\hline
 Frequency (Ghz) & $A$ (mJy) & $B$ (mJy) &  $\phi_0$ & $\chi^2$ \\
4.85   & $9.3 \pm 1.2$ & $-0.2 \pm 0.8$ & $0.7 \pm 0.1$ &  0.9  \\ 
8.35 & $8.9 \pm 0.9$ & $2.1 \pm 0.8$  & $0.7 \pm 0.1$ & 0.8 \\
10.45 & $6.7 \pm 0.7$ & $0.5 \pm 0.6$ &  $0.7 \pm 0.1$  & 1.3 \\ \hline
\label{table:paramsine}
\end{tabular}
\end{table}


\begin{figure}
\begin{center}
\includegraphics[width=.6\columnwidth]{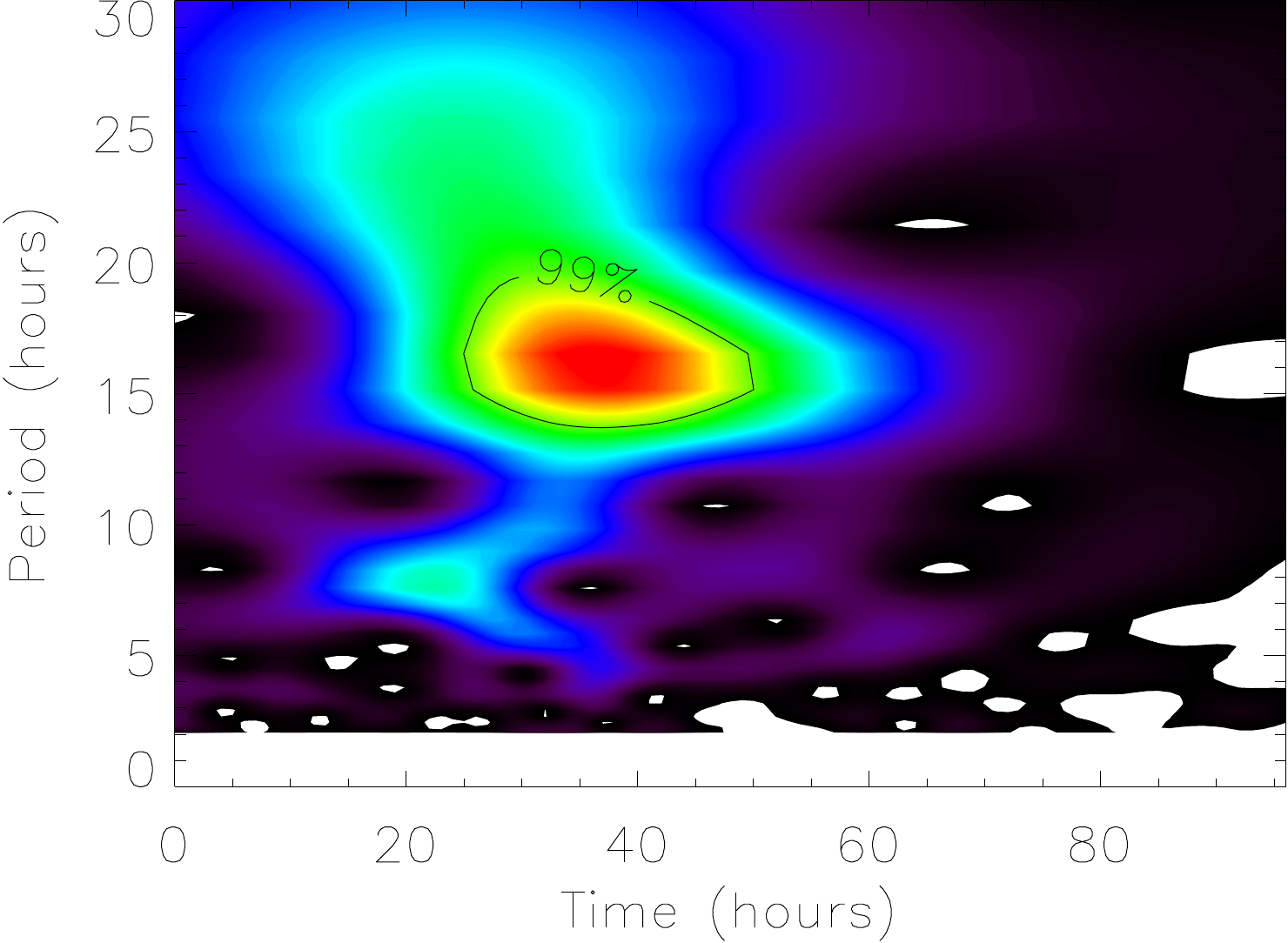}\\
\includegraphics[width=.6\columnwidth]{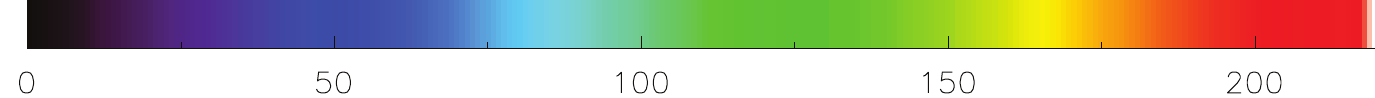}\\
\includegraphics[width=.49\columnwidth]{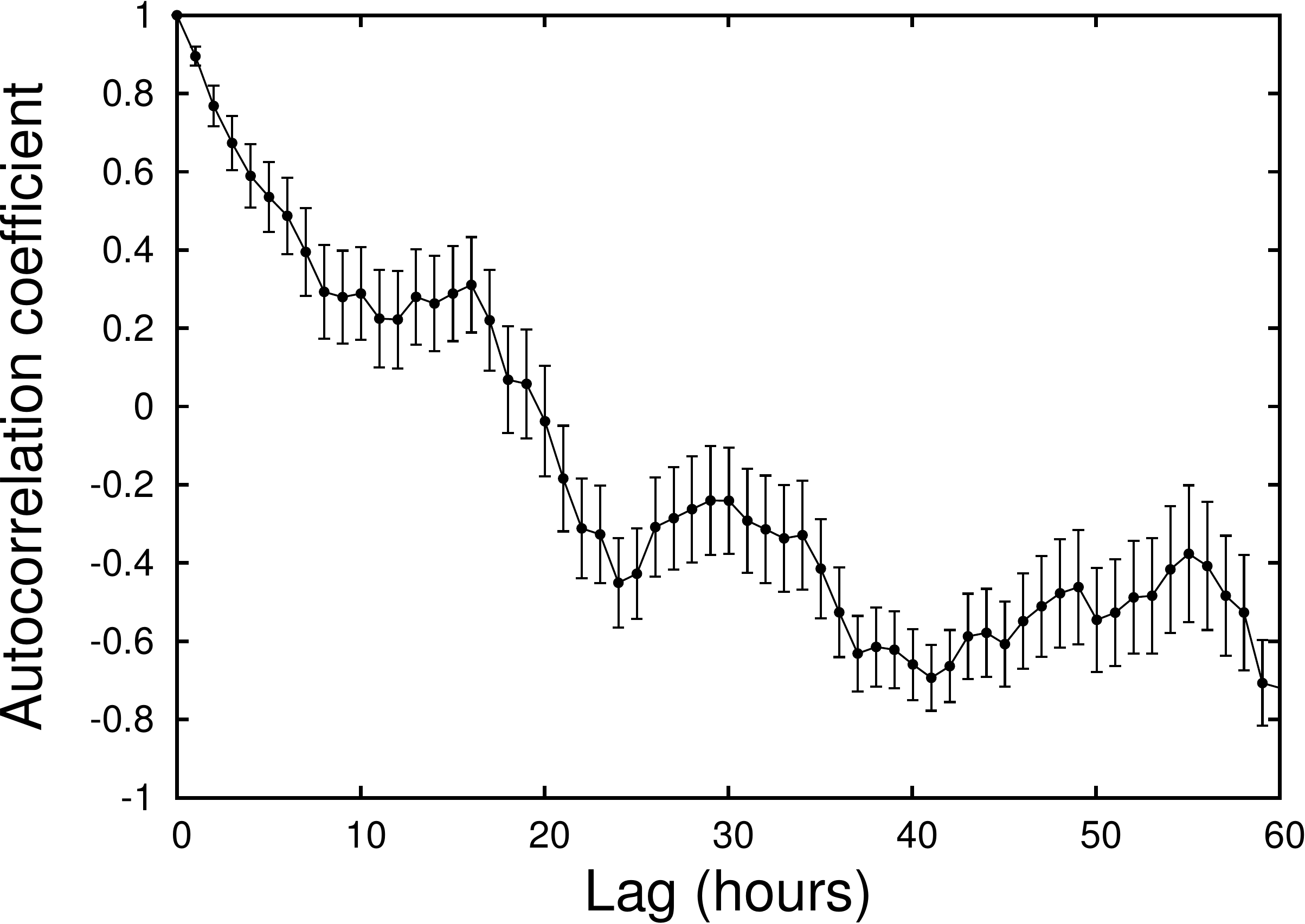}
\includegraphics[width=.49\columnwidth]{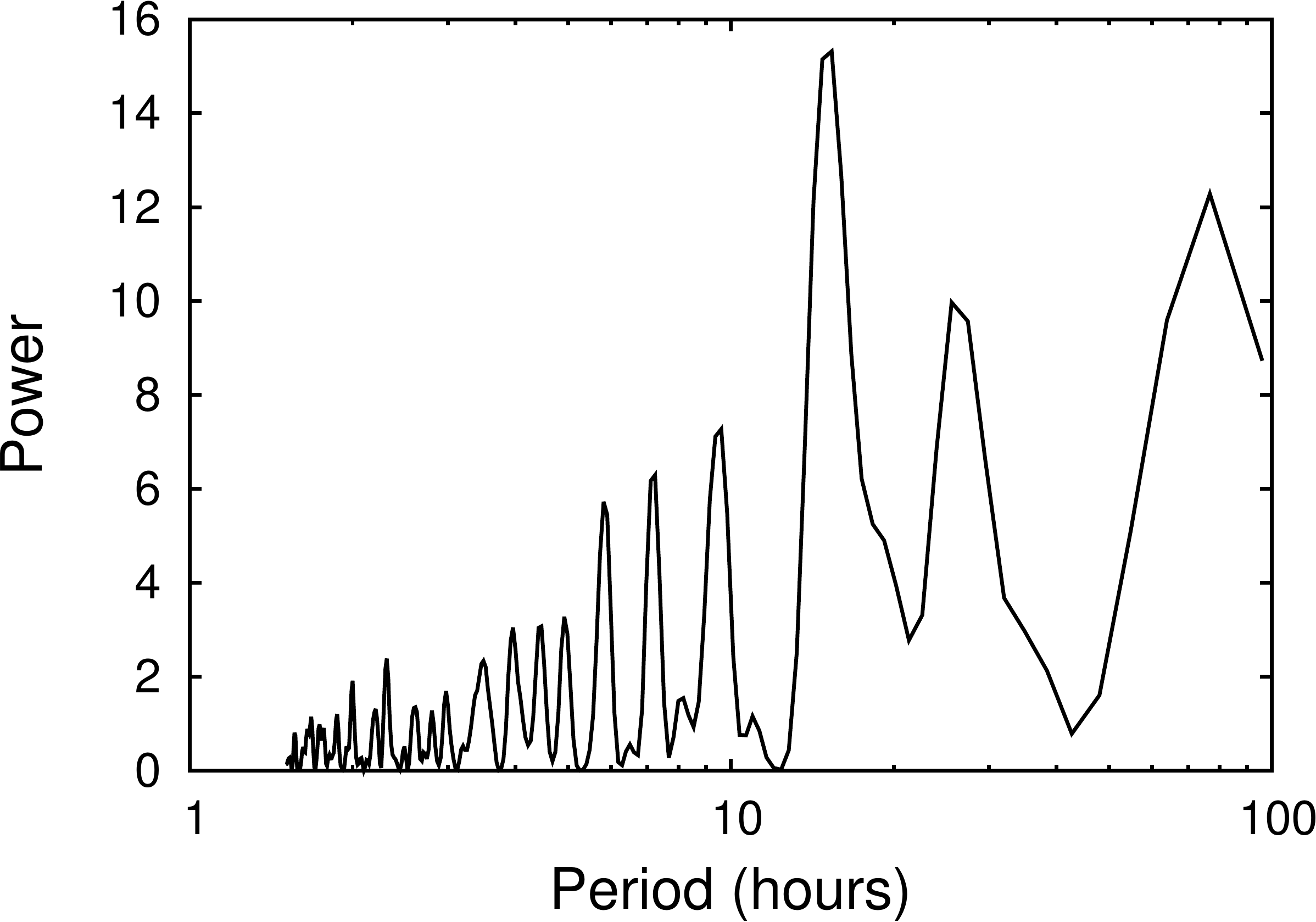} 
\caption{Timing analysis results of the 4.85 GHz data. Top: Wavelet plot with maximum power (red) around 16 hours, at the 99 per cent confidence level. Bottom left: Auto-correlation function. Bottom right: Lomb-Scargle periodogram with most significant peak at 15.4 hours.}
\label{fig:timinganalysis}
\end{center}
\end{figure}

\begin{figure*}
\begin{center}
\includegraphics[width=0.65\columnwidth]{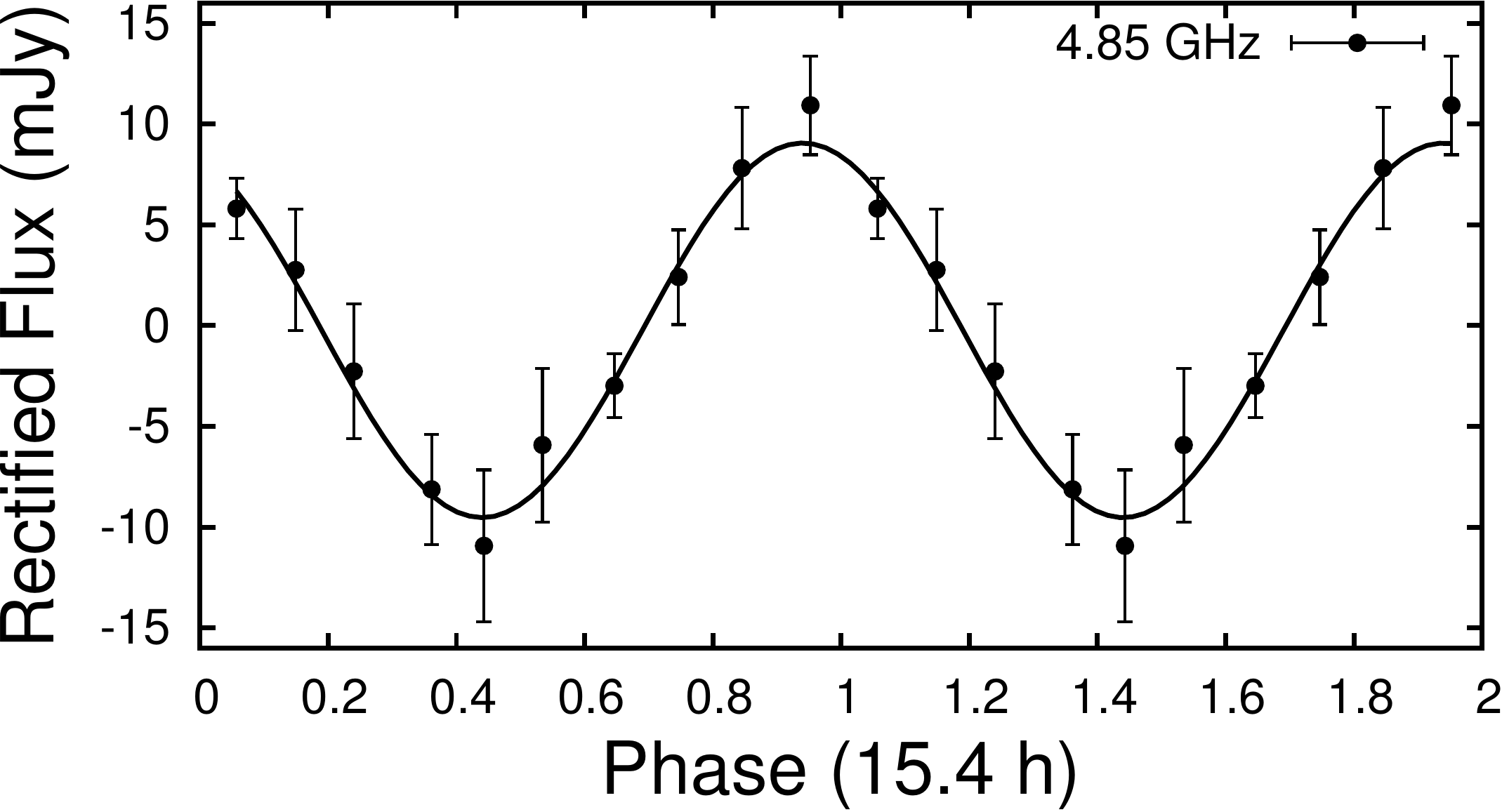}
\includegraphics[width=0.65\columnwidth]{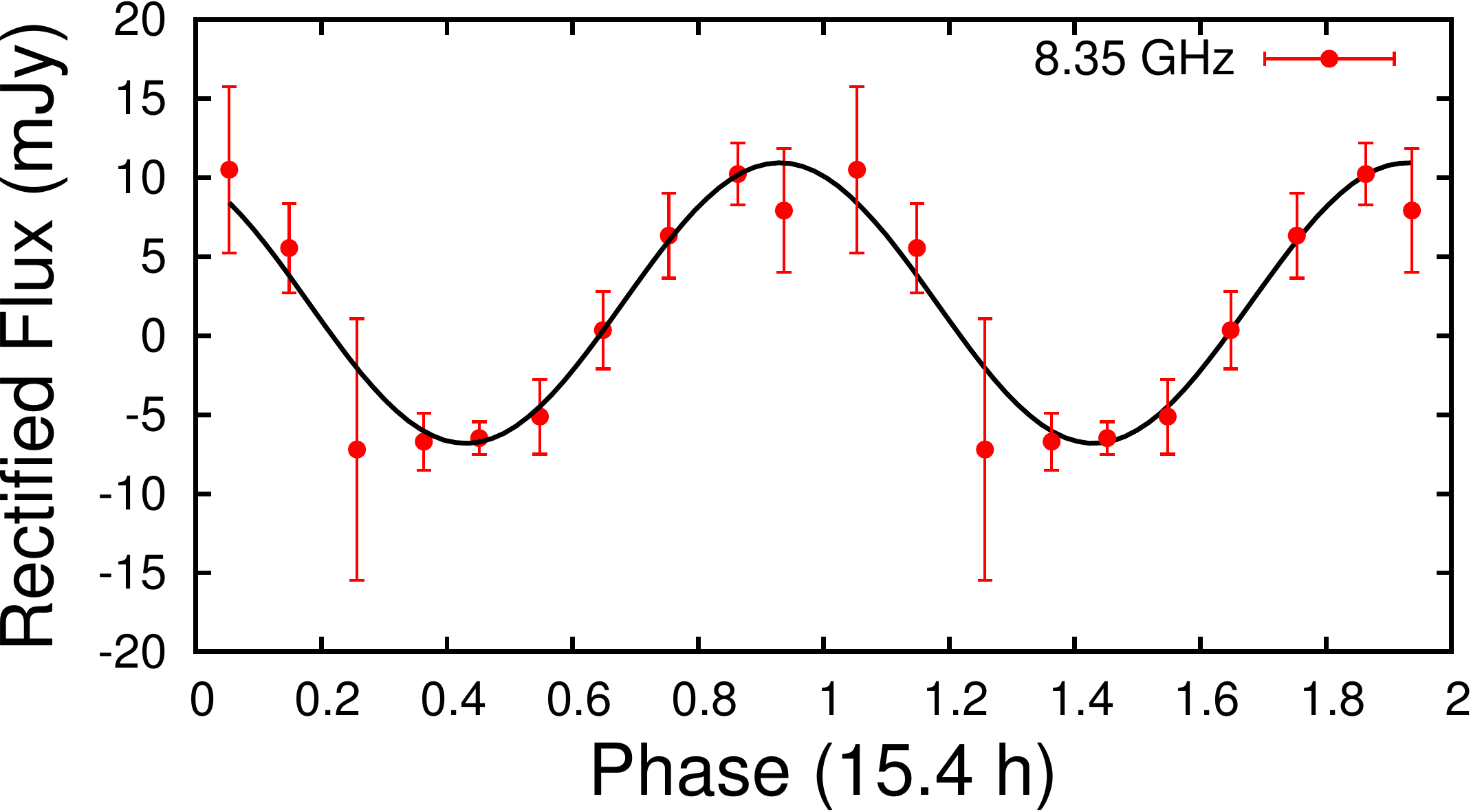}
\includegraphics[width=0.65\columnwidth]{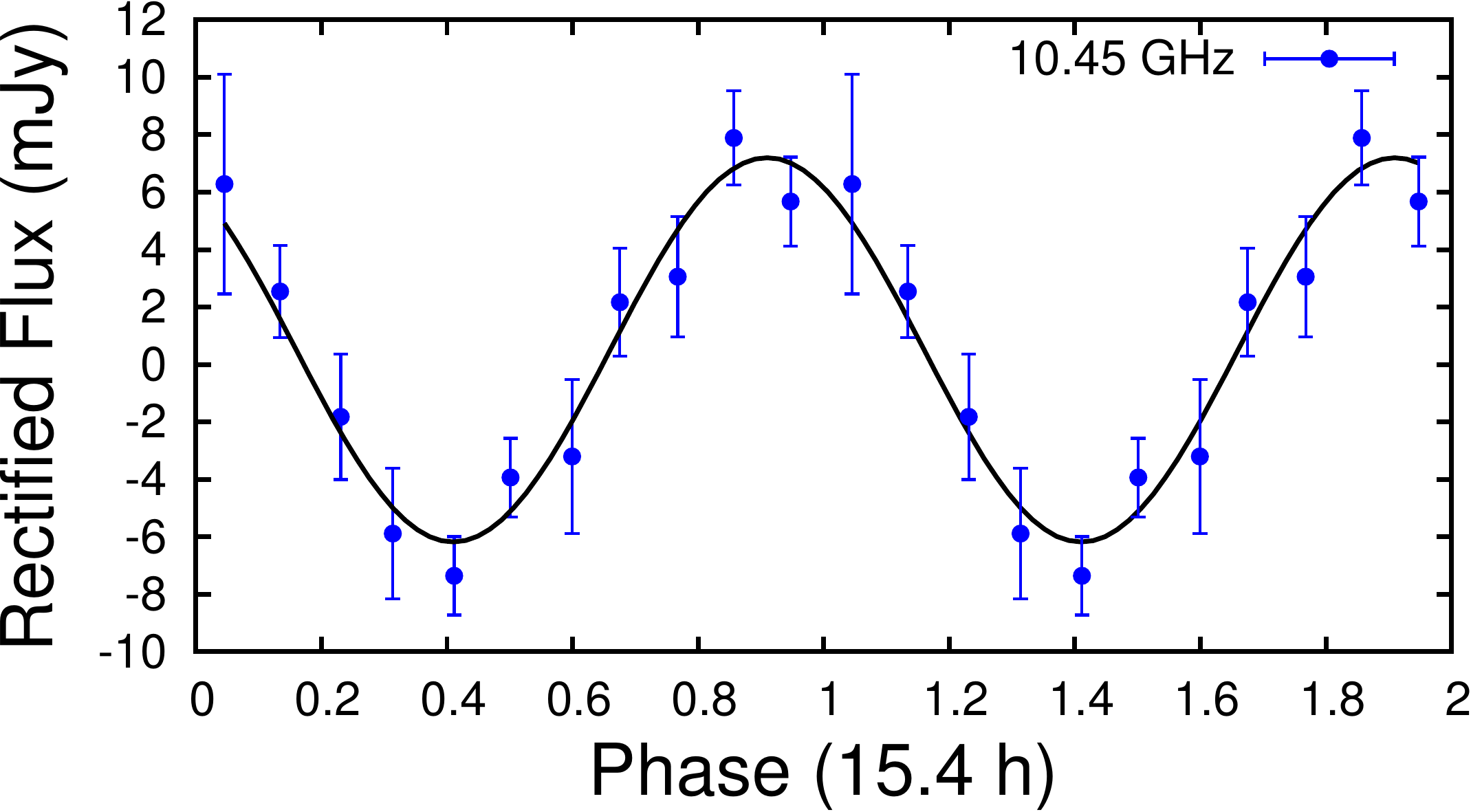}
\caption{Rectified flux folded on the period of 15.4 h, which resulted most significant from timing analysis. Error bars reflect the $1\sigma$ uncertainties resulting from the standard deviation in each phase bin. The data are repeated in the second cycle. The solid line (black) is a result of fitting a sine function to the folded data. The data are taken from the time-intervals:
4.85~GHz,     MJD 56765.64 -- 56768.48 (4.4~cycles);
8.35~GHz,     MJD 56766.06 -- 56767.64 (2.5~cycles);
10.45~GHz,    MJD 56765.98 -- 56768.48 (3.9~cycles).}
\label{fig:fold}
\end{center}
\end{figure*}

The light curves at all three frequencies (Fig.~\ref{fig:lightcurves}\,a, c, e) show small-scale oscillations. Amplitude and width seem to vary from one peak to the other.
The higher sensitivity of 4.85~GHz data give rise to significant timing analysis results: the wavelet analysis in the top panel of Fig.~\ref{fig:timinganalysis} shows that the oscillations at 4.85~GHz have a periodicity of about 16~hours.  The auto-correlation coefficient in the middle panel of Fig.~\ref{fig:timinganalysis} shows peaks at 15~hours, 30~hours and at about 55~hours. Lomb-Scargle analysis in the bottom panel of Figure~\ref{fig:timinganalysis} gives    
a dominant and significant feature at $15.4 \pm 0.6$\,h. Indeed, data at all three frequencies fold with the 15.4~hours period (see Fig.~\ref{fig:fold} and fitting parameters in Table~\ref{table:paramsine}) with significance of the oscillations clearly above $8\sigma$.
Oscillations are also present in the radio spectral index $\alpha$ (Fig.~\ref{fig:lightcurves}\,d)
corroborating the observed spectral index oscillations  
by \citet[][see the bottom panel of their Fig.~1]{Zimmermann2015}.
The oscillations create a sort of flattening of the spectral index.
and a zoom-in centered at MJD 56766.3 clearly shows $\alpha = 0$ (Fig.~\ref{fig:lightcurves}\,f). 
%

\section{Conclusions and Discussion}

We observed one radio outburst of \lsi{} with the Effelsberg 100-m telescope in April 2014 for approximately 100~h at 4.85, 8.35, and 10.45 GHz. We analysed these data along with simultaneous \fermilat{} GeV data. Our results reveal the following:
\begin{enumerate}
\item{
QPO previously observed in \lsi{} showed time scales of minutes, hours and days \citep{Taylor1992,Peracaula1997,Zimmermann2015}. Our study determines periodicities of $\sim 15$\,h. 
There are three hypotheses to explain the physical mechanism behind the occurrence of these periodicities. The first is associated to the geometry of the jet \citep{Rani2010}, the second one to multiple shocks \citep[][and references therein]{Klein-Wolt2002}, and the third one implies discrete plasma ejections \citep[][and references therein]{Mirabel1999}.
In the first scenario the reason for oscillations is the helical topology of the magnetic field in the jet associated to Doppler boosting effects. 
Since in a conical jet of a microquasar the 10 GHz emission originates from a jet-segment nearer to the central engine of the system than the 5~GHz jet-segment 
 \citep{kaiser2006}, this scenario  implies a  longer period for 5 GHz oscillations.
This is not compatible with our results in Table~\ref{table:paramsine} and Fig.~\ref{fig:fold}, 
giving the same period for the oscillations at all three observed frequencies.
Following the \citet{Vanderlaan1966} model of adiabatically expanding spheroidal ejecta
of plasma, the low-frequency emission peak is delayed and weaker than that at higher frequencies.
Our result of a larger amplitude at lower frequency is in contradiction with this model.
Finally, the shock-in-jet model proposed by \citet{Marschergear1985} and generalised by \citet{Valtaoja1992} predicts 
flare amplitudes to increase towards lower frequency, as observed.}
\item{
As previously observed  in X-ray binaries \grs{} and \cgx{}, the radio spectral index of \lsi{} oscillates in phase with total flux variations,
the radio spectral index modulation being superimposed on a longer modulation whose peak corresponds to the flattening of the spectral index. This finding is consistent with the hypothesis that the ``flat'' radio spectra in microquasars arise from the combination of emission from optically thick and thin regions \citep{Fender2002}.
The radio spectrum after the oscillations becomes optically thin with $p \sim 2.0$, the $\gamma$-ray emission follows a similar trend with a photon index~$\Gamma \sim 2.0$.}
\end{enumerate}


\section*{Acknowledgements}

We thank Eduardo Ros for carefully reading the manuscript. This work has made use of public \textit{Fermi} data obtained from the High Energy Astrophysics Science Archive Research Center (HEASARC), provided by NASA Goddard Space Flight Center.




\bibliographystyle{mnras}
\bibliography{reference} 

\begin{thebibliography}{}
\makeatletter
\relax
\def\mn@urlcharsother{\let\do\@makeother \do\$\do\&\do\#\do\^\do\_\do\%\do\~}
\def\mn@doi{\begingroup\mn@urlcharsother \@ifnextchar [ {\mn@doi@}
  {\mn@doi@[]}}
\def\mn@doi@[#1]#2{\def\@tempa{#1}\ifx\@tempa\@empty \href
  {http://dx.doi.org/#2} {doi:#2}\else \href {http://dx.doi.org/#2} {#1}\fi
  \endgroup}
\def\mn@eprint#1#2{\mn@eprint@#1:#2::\@nil}
\def\mn@eprint@arXiv#1{\href {http://arxiv.org/abs/#1} {{\tt arXiv:#1}}}
\def\mn@eprint@dblp#1{\href {http://dblp.uni-trier.de/rec/bibtex/#1.xml}
  {dblp:#1}}
\def\mn@eprint@#1:#2:#3:#4\@nil{\def\@tempa {#1}\def\@tempb {#2}\def\@tempc
  {#3}\ifx \@tempc \@empty \let \@tempc \@tempb \let \@tempb \@tempa \fi \ifx
  \@tempb \@empty \def\@tempb {arXiv}\fi \@ifundefined
  {mn@eprint@\@tempb}{\@tempb:\@tempc}{\expandafter \expandafter \csname
  mn@eprint@\@tempb\endcsname \expandafter{\@tempc}}}

\bibitem[\protect\citeauthoryear{{Abdo} et~al.,}{{Abdo}
  et~al.}{2009}]{Abdo2009}
{Abdo} A.~A.,  et~al., 2009, \mn@doi [\apjl] {10.1088/0004-637X/701/2/L123},
  \href {http://adsabs.harvard.edu/abs/2009ApJ...701L.123A} {701, L123}

\bibitem[\protect\citeauthoryear{{Acciari} et~al.,}{{Acciari}
  et~al.}{2008}]{Acciari2008}
{Acciari} V.~A.,  et~al., 2008, \mn@doi [\apj] {10.1086/587736}, \href
  {http://adsabs.harvard.edu/abs/2008ApJ...679.1427A} {679, 1427}

\bibitem[\protect\citeauthoryear{{Acciari} et~al.,}{{Acciari}
  et~al.}{2011}]{Acciari2011}
{Acciari} V.~A.,  et~al., 2011, \mn@doi [\apj] {10.1088/0004-637X/738/1/3},
  738, 3

\bibitem[\protect\citeauthoryear{{Acero} et~al.,}{{Acero}
  et~al.}{2015}]{FermiLAT2015}
{Acero} F.,  et~al., 2015, \mn@doi [\apjs] {10.1088/0067-0049/218/2/23}, \href
  {http://adsabs.harvard.edu/abs/2015ApJS..218...23A} {218, 23}

\bibitem[\protect\citeauthoryear{{Ackermann} et~al.,}{{Ackermann}
  et~al.}{2013}]{Ackermann2013}
{Ackermann} M.,  et~al., 2013, \mn@doi [\apjl] {10.1088/2041-8205/773/2/L35},
  \href {http://adsabs.harvard.edu/abs/2013ApJ...773L..35A} {773, L35}

\bibitem[\protect\citeauthoryear{{Ahnen} et~al.,}{{Ahnen}
  et~al.}{2016}]{Ahnen2016}
{Ahnen} M.~L.,  et~al., 2016, \mn@doi [\aap] {10.1051/0004-6361/201527964},
  \href {http://adsabs.harvard.edu/abs/2016A%26A...591A..76A} {591, A76}

\bibitem[\protect\citeauthoryear{{Albert} et~al.,}{{Albert}
  et~al.}{2006}]{Albert2006}
{Albert} J.,  et~al., 2006, \mn@doi [Science] {10.1126/science.1128177}, \href
  {http://adsabs.harvard.edu/abs/2006Sci...312.1771A} {312, 1771}

\bibitem[\protect\citeauthoryear{{Angelakis} et~al.,}{{Angelakis}
  et~al.}{2015}]{Angelakis2015}
{Angelakis} E.,  et~al., 2015, \mn@doi [\aap] {10.1051/0004-6361/201425081},
  \href {http://adsabs.harvard.edu/abs/2015A%26A...575A..55A} {575, A55}

\bibitem[\protect\citeauthoryear{{Casares}, {Ribas}, {Paredes}, {Mart{\'{\i}}}
  \& {Allende Prieto}}{{Casares} et~al.}{2005}]{Casares2005}
{Casares} J.,  {Ribas} I.,  {Paredes} J.~M.,  {Mart{\'{\i}}} J.,   {Allende
  Prieto} C.,  2005, \mn@doi [\mnras] {10.1111/j.1365-2966.2005.09106.x}, \href
  {http://adsabs.harvard.edu/abs/2005MNRAS.360.1105C} {360, 1105}

\bibitem[\protect\citeauthoryear{{Dubus}}{{Dubus}}{2013}]{Dubus2013}
{Dubus} G.,  2013, \mn@doi [\aapr] {10.1007/s00159-013-0064-5}, \href
  {http://adsabs.harvard.edu/abs/2013A%26ARv..21...64D} {21, 64}

\bibitem[\protect\citeauthoryear{{Espaillat}, {Bregman}, {Hughes}  \&
  {Lloyd-Davies}}{{Espaillat} et~al.}{2008}]{espaillat2008}
{Espaillat} C.,  {Bregman} J.,  {Hughes} P.,   {Lloyd-Davies} E.,  2008,
  \mn@doi [\apj] {10.1086/587023}, \href
  {http://adsabs.harvard.edu/abs/2008ApJ...679..182E} {679, 182}

\bibitem[\protect\citeauthoryear{{Fender} \& {Pooley}}{{Fender} \&
  {Pooley}}{1998}]{Fender1998}
{Fender} R.~P.,  {Pooley} G.~G.,  1998, \mn@doi [\mnras]
  {10.1046/j.1365-8711.1998.01921.x}, \href
  {http://adsabs.harvard.edu/abs/1998MNRAS.300..573F} {300, 573}

\bibitem[\protect\citeauthoryear{{Fender}, {Pooley}, {Robinson}, {Harmon},
  {Zhang}  \& {Canosa}}{{Fender} et~al.}{1997}]{fender97}
{Fender} R.~P.,  {Pooley} G.~G.,  {Robinson} C.~R.,  {Harmon} B.~A.,  {Zhang}
  S.~N.,   {Canosa} C.,  1997, in {Wickramasinghe} D.~T.,  {Bicknell} G.~V.,
  {Ferrario} L.,  eds,  Astronomical Society of the Pacific Conference Series
  Vol. 121, IAU Colloq. 163: Accretion Phenomena and Related Outflows. p.~701
  (\mn@eprint {} {astro-ph/9612092})

\bibitem[\protect\citeauthoryear{{Fender}, {Rayner}, {Trushkin}, {O'Brien},
  {Sault}, {Pooley}  \& {Norris}}{{Fender} et~al.}{2002}]{Fender2002}
{Fender} R.~P.,  {Rayner} D.,  {Trushkin} S.~A.,  {O'Brien} K.,  {Sault} R.~J.,
   {Pooley} G.~G.,   {Norris} R.~P.,  2002, \mn@doi [\mnras]
  {10.1046/j.1365-8711.2002.05072.x}, \href
  {http://adsabs.harvard.edu/abs/2002MNRAS.330..212F} {330, 212}

\bibitem[\protect\citeauthoryear{{Gierli{\'n}ski}, {Middleton}, {Ward}  \&
  {Done}}{{Gierli{\'n}ski} et~al.}{2008}]{Gierlinski2008}
{Gierli{\'n}ski} M.,  {Middleton} M.,  {Ward} M.,   {Done} C.,  2008, \mn@doi
  [\nat] {10.1038/nature07277}, \href
  {http://adsabs.harvard.edu/abs/2008Natur.455..369G} {455, 369}

\bibitem[\protect\citeauthoryear{{Gregory}}{{Gregory}}{2002}]{Gregory2002}
{Gregory} P.~C.,  2002, \mn@doi [\apj] {10.1086/341257}, \href
  {http://adsabs.harvard.edu/abs/2002ApJ...575..427G} {575, 427}

\bibitem[\protect\citeauthoryear{{Hadasch} et~al.,}{{Hadasch}
  et~al.}{2012}]{Hadasch2012}
{Hadasch} D.,  et~al., 2012, \mn@doi [\apj] {10.1088/0004-637X/749/1/54}, \href
  {http://adsabs.harvard.edu/abs/2012ApJ...749...54H} {749, 54}

\bibitem[\protect\citeauthoryear{{Han} \& {Hjellming}}{{Han} \&
  {Hjellming}}{1992}]{Han1992}
{Han} X.,  {Hjellming} R.~M.,  1992, \mn@doi [\apj] {10.1086/171996}, \href
  {http://adsabs.harvard.edu/abs/1992ApJ...400..304H} {400, 304}

\bibitem[\protect\citeauthoryear{{Harrison}, {Ray}, {Leahy}, {Waltman}  \&
  {Pooley}}{{Harrison} et~al.}{2000}]{Harrison2000}
{Harrison} F.~A.,  {Ray} P.~S.,  {Leahy} D.~A.,  {Waltman} E.~B.,   {Pooley}
  G.~G.,  2000, \mn@doi [\apj] {10.1086/308157}, \href
  {http://adsabs.harvard.edu/abs/2000ApJ...528..454H} {528, 454}

\bibitem[\protect\citeauthoryear{{Jaron} \& {Massi}}{{Jaron} \&
  {Massi}}{2013}]{Jaron2013}
{Jaron} F.,  {Massi} M.,  2013, \mn@doi [\aap] {10.1051/0004-6361/201322557},
  \href {http://adsabs.harvard.edu/abs/2013A%26A...559A.129J} {559, A129}

\bibitem[\protect\citeauthoryear{{Jaron}, {Torricelli-Ciamponi}  \&
  {Massi}}{{Jaron} et~al.}{2016}]{Jaron2016}
{Jaron} F.,  {Torricelli-Ciamponi} G.,   {Massi} M.,  2016, \mn@doi [\aap]
  {10.1051/0004-6361/201628556}, \href
  {http://adsabs.harvard.edu/abs/2016A%26A...595A..92J} {595, A92}

\bibitem[\protect\citeauthoryear{{Kaiser}}{{Kaiser}}{2006}]{kaiser2006}
{Kaiser} C.~R.,  2006, \mn@doi [\mnras] {10.1111/j.1365-2966.2006.10030.x},
  \href {http://adsabs.harvard.edu/abs/2006MNRAS.367.1083K} {367, 1083}

\bibitem[\protect\citeauthoryear{{Klein-Wolt}, {Fender}, {Pooley}, {Belloni},
  {Migliari}, {Morgan}  \& {van der Klis}}{{Klein-Wolt}
  et~al.}{2002}]{Klein-Wolt2002}
{Klein-Wolt} M.,  {Fender} R.~P.,  {Pooley} G.~G.,  {Belloni} T.,  {Migliari}
  S.,  {Morgan} E.~H.,   {van der Klis} M.,  2002, \mn@doi [\mnras]
  {10.1046/j.1365-8711.2002.05223.x}, \href
  {http://adsabs.harvard.edu/abs/2002MNRAS.331..745K} {331, 745}

\bibitem[\protect\citeauthoryear{{Li}, {Torres}  \& {Zhang}}{{Li}
  et~al.}{2014}]{Li2014}
{Li} J.,  {Torres} D.~F.,   {Zhang} S.,  2014, \mn@doi [\apjl]
  {10.1088/2041-8205/785/1/L19}, \href
  {http://adsabs.harvard.edu/abs/2014ApJ...785L..19L} {785, L19}

\bibitem[\protect\citeauthoryear{{Linnell Nemec} \& {Nemec}}{{Linnell Nemec} \&
  {Nemec}}{1985}]{Nemec1985}
{Linnell Nemec} A.~F.,  {Nemec} J.~M.,  1985, \mn@doi [\aj] {10.1086/113936},
  \href {http://adsabs.harvard.edu/abs/1985AJ.....90.2317L} {90, 2317}

\bibitem[\protect\citeauthoryear{{Lomb}}{{Lomb}}{1976}]{Lomb1976}
{Lomb} N.~R.,  1976, \mn@doi [\apss] {10.1007/BF00648343}, \href
  {http://adsabs.harvard.edu/abs/1976Ap%26SS..39..447L} {39, 447}

\bibitem[\protect\citeauthoryear{{Marscher} \& {Gear}}{{Marscher} \&
  {Gear}}{1985}]{Marschergear1985}
{Marscher} A.~P.,  {Gear} W.~K.,  1985, \mn@doi [\apj] {10.1086/163592}, \href
  {http://adsabs.harvard.edu/abs/1985ApJ...298..114M} {298, 114}

\bibitem[\protect\citeauthoryear{{Marti}, {Mirabel}  \& {Rodriguez}}{{Marti}
  et~al.}{2001}]{Marti2001}
{Marti} J.,  {Mirabel} I.~F.,   {Rodriguez} L.~F.,  2001, Information Bulletin
  on Variable Stars, \href {http://adsabs.harvard.edu/abs/2001IBVS.5127....1M}
  {5127}

\bibitem[\protect\citeauthoryear{{Massi} \& {Torricelli-Ciamponi}}{{Massi} \&
  {Torricelli-Ciamponi}}{2014}]{Massi2014}
{Massi} M.,  {Torricelli-Ciamponi} G.,  2014, \mn@doi [\aap]
  {10.1051/0004-6361/201322760}, \href
  {http://adsabs.harvard.edu/abs/2014A%26A...564A..23M} {564, A23}

\bibitem[\protect\citeauthoryear{{Massi} \& {Torricelli-Ciamponi}}{{Massi} \&
  {Torricelli-Ciamponi}}{2016}]{MassiTorricelli2016}
{Massi} M.,  {Torricelli-Ciamponi} G.,  2016, \mn@doi [\aap]
  {10.1051/0004-6361/201526938}, \href
  {http://adsabs.harvard.edu/abs/2016A%26A...585A.123M} {585, A123}

\bibitem[\protect\citeauthoryear{{Massi}, {Migliari}  \& {Chernyakova}}{{Massi}
  et~al.}{2017}]{Massi2017}
{Massi} M.,  {Migliari} S.,   {Chernyakova} M.,  2017, preprint, \href
  {http://adsabs.harvard.edu/abs/2017arXiv170401335M} {} (\mn@eprint {arXiv}
  {1704.01335})

\bibitem[\protect\citeauthoryear{{Mirabel} \& {Rodr{\'{\i}}guez}}{{Mirabel} \&
  {Rodr{\'{\i}}guez}}{1999}]{Mirabel1999}
{Mirabel} I.~F.,  {Rodr{\'{\i}}guez} L.~F.,  1999, \mn@doi [\araa]
  {10.1146/annurev.astro.37.1.409}, \href
  {http://adsabs.harvard.edu/abs/1999ARA%26A..37..409M} {37, 409}

\bibitem[\protect\citeauthoryear{{Nagae}, {Kawabata}, {Fukazawa}, {Okazaki}  \&
  {Isogai}}{{Nagae} et~al.}{2008}]{Nagae2008}
{Nagae} O.,  {Kawabata} K.~S.,  {Fukazawa} Y.,  {Okazaki} A.,   {Isogai} M.,
  2008, in {Yuan} Y.-F.,  {Li} X.-D.,   {Lai} D.,  eds,  American Institute of
  Physics Conference Series Vol. 968, Astrophysics of Compact Objects. pp
  328--330, \mn@doi{10.1063/1.2840421}

\bibitem[\protect\citeauthoryear{{Paredes-Fortuny}, {Rib{\'o}}, {Bosch-Ramon},
  {Casares}, {Fors}  \& {N{\'u}{\~n}ez}}{{Paredes-Fortuny}
  et~al.}{2015}]{Paredes-Fortuny2015}
{Paredes-Fortuny} X.,  {Rib{\'o}} M.,  {Bosch-Ramon} V.,  {Casares} J.,  {Fors}
  O.,   {N{\'u}{\~n}ez} J.,  2015, \mn@doi [\aap]
  {10.1051/0004-6361/201425361}, \href
  {http://adsabs.harvard.edu/abs/2015A%26A...575L...6P} {575, L6}

\bibitem[\protect\citeauthoryear{{Peracaula}, {Marti}  \&
  {Paredes}}{{Peracaula} et~al.}{1997}]{Peracaula1997}
{Peracaula} M.,  {Marti} J.,   {Paredes} J.~M.,  1997, \aap, \href
  {http://adsabs.harvard.edu/abs/1997A%26A...328..283P} {328, 283}

\bibitem[\protect\citeauthoryear{{Plotkin} et~al.,}{{Plotkin}
  et~al.}{2017}]{Plotkin2017}
{Plotkin} R.~M.,  et~al., 2017, \mn@doi [\apj] {10.3847/1538-4357/834/2/104},
  \href {http://adsabs.harvard.edu/abs/2017ApJ...834..104P} {834, 104}

\bibitem[\protect\citeauthoryear{{Pooley} \& {Fender}}{{Pooley} \&
  {Fender}}{1997}]{Pooley1997}
{Pooley} G.~G.,  {Fender} R.~P.,  1997, \mn@doi [\mnras]
  {10.1093/mnras/292.4.925}, \href
  {http://adsabs.harvard.edu/abs/1997MNRAS.292..925P} {292, 925}

\bibitem[\protect\citeauthoryear{{Rani}, {Gupta}, {Joshi}, {Ganesh}  \&
  {Wiita}}{{Rani} et~al.}{2010}]{Rani2010}
{Rani} B.,  {Gupta} A.~C.,  {Joshi} U.~C.,  {Ganesh} S.,   {Wiita} P.~J.,
  2010, \mn@doi [\apjl] {10.1088/2041-8205/719/2/L153}, \href
  {http://adsabs.harvard.edu/abs/2010ApJ...719L.153R} {719, L153}

\bibitem[\protect\citeauthoryear{{Rea} \& {Torres}}{{Rea} \&
  {Torres}}{2008}]{Rea2008}
{Rea} N.,  {Torres} D.~F.,  2008, The Astronomer's Telegram, \href
  {http://adsabs.harvard.edu/abs/2008ATel.1731....1R} {1731}

\bibitem[\protect\citeauthoryear{{Rivinius}, {Carciofi}  \&
  {Martayan}}{{Rivinius} et~al.}{2013}]{Rivinius2013}
{Rivinius} T.,  {Carciofi} A.~C.,   {Martayan} C.,  2013, \mn@doi [\aapr]
  {10.1007/s00159-013-0069-0}, \href
  {http://adsabs.harvard.edu/abs/2013A%26ARv..21...69R} {21, 69}

\bibitem[\protect\citeauthoryear{{Rodr{\'{\i}}guez} \&
  {Mirabel}}{{Rodr{\'{\i}}guez} \& {Mirabel}}{1997}]{Rodriguez1997}
{Rodr{\'{\i}}guez} L.~F.,  {Mirabel} I.~F.,  1997, \mn@doi [\apjl]
  {10.1086/310443}, \href {http://adsabs.harvard.edu/abs/1997ApJ...474L.123R}
  {474, L123}

\bibitem[\protect\citeauthoryear{{Scargle}}{{Scargle}}{1982}]{Scargle1982}
{Scargle} J.~D.,  1982, \mn@doi [\apj] {10.1086/160554}, \href
  {http://adsabs.harvard.edu/abs/1982ApJ...263..835S} {263, 835}

\bibitem[\protect\citeauthoryear{{Taylor}, {Kenny}, {Spencer}  \&
  {Tzioumis}}{{Taylor} et~al.}{1992}]{Taylor1992}
{Taylor} A.~R.,  {Kenny} H.~T.,  {Spencer} R.~E.,   {Tzioumis} A.,  1992,
  \mn@doi [\apj] {10.1086/171648}, \href
  {http://adsabs.harvard.edu/abs/1992ApJ...395..268T} {395, 268}

\bibitem[\protect\citeauthoryear{{Torrence} \& {Compo}}{{Torrence} \&
  {Compo}}{1998}]{Torrence1998}
{Torrence} C.,  {Compo} G.~P.,  1998, \mn@doi [Bulletin of the American
  Meteorological Society] {10.1175/1520-0477(1998)079<0061:APGTWA>2.0.CO;2},
  \href {http://adsabs.harvard.edu/abs/1998BAMS...79...61T} {79, 61}

\bibitem[\protect\citeauthoryear{{Torres}, {Rea}, {Esposito}, {Li}, {Chen}  \&
  {Zhang}}{{Torres} et~al.}{2012}]{Torres2012}
{Torres} D.~F.,  {Rea} N.,  {Esposito} P.,  {Li} J.,  {Chen} Y.,   {Zhang} S.,
  2012, \mn@doi [\apj] {10.1088/0004-637X/744/2/106}, \href
  {http://adsabs.harvard.edu/abs/2012ApJ...744..106T} {744, 106}

\bibitem[\protect\citeauthoryear{{Valtaoja}, {Terasranta}, {Urpo}, {Nesterov},
  {Lainela}  \& {Valtonen}}{{Valtaoja} et~al.}{1992}]{Valtaoja1992}
{Valtaoja} E.,  {Terasranta} H.,  {Urpo} S.,  {Nesterov} N.~S.,  {Lainela} M.,
   {Valtonen} M.,  1992, \aap, \href
  {http://adsabs.harvard.edu/abs/1992A%26A...254...71V} {254, 71}

\bibitem[\protect\citeauthoryear{{Van der Laan}}{{Van der
  Laan}}{1966}]{Vanderlaan1966}
{Van der Laan} H.,  1966, \mn@doi [\nat] {10.1038/2111131a0}, \href
  {http://adsabs.harvard.edu/abs/1966Natur.211.1131V} {211, 1131}

\bibitem[\protect\citeauthoryear{{Zimmermann}, {Fuhrmann}  \&
  {Massi}}{{Zimmermann} et~al.}{2015}]{Zimmermann2015}
{Zimmermann} L.,  {Fuhrmann} L.,   {Massi} M.,  2015, \mn@doi [\aap]
  {10.1051/0004-6361/201425545}, \href
  {http://adsabs.harvard.edu/abs/2015A%26A...580L...2Z} {580, L2}

\makeatother
\end{thebibliography}
\bsp	
\label{lastpage}
\end{document}